# Global Diffusion in a Realistic Three-Dimensional Time-Dependent Nonturbulent Fluid Flow

Julyan H. E. Cartwright[1,*], Mario Feingold[2,†], and Oreste Piro[1,3,‡]

[1]*Departament de Física, Universitat de les Illes Balears, 07071 Palma de Mallorca, Spain*
[2]*Department of Physics, Ben-Gurion University, Beer-Sheva 84105, Israel*
[3]*Institut Mediterrani d'Estudis Avançats (CSIC–UIB), 07071 Palma de Mallorca, Spain*

We introduce and study the first model of an experimentally realizable three-dimensional time-dependent nonturbulent fluid flow to display the phenomenon of global diffusion of passive-scalar particles at arbitrarily small values of the nonintegrable perturbation. This type of chaotic advection, termed *resonance-induced diffusion*, is generic for a large class of flows.

Dynamical systems that arise in problems of particle diffusion [1] in incompressible and nonturbulent fluid flows, apart from being of theoretical interest, hold much relevance for technological applications. Properties of emulsions, dispersion of contaminants in the atmosphere and ocean, sedimentation, and mixing, are just a few examples. In comparison with the present state of knowledge of chaotic advection, mixing and transport in two-dimensional [2] or three-dimensional time-independent [3] flows, little is known for three-dimensional time-dependent flows.

The strobed dynamics of fluid parcels — so called passive scalars — in three-dimensional time-dependent incompressible fluid flow is qualitatively equivalent to the iteration of a three-dimensional volume-preserving map. A few years ago we began to investigate nearly-integrable classes of such maps — which we have termed *Liouvillian* maps — in search of possible generic features [4]. We found that we could characterize Liouvillian maps in a similar manner to Hamiltonian dynamical systems, by using the number of slow action and fast angle variables that, in the integrable limit, remain invariant and rotate uniformly, respectively. The cases having one or two of such actions were found to be particularly interesting. For the one-action case, which has been shown to be generic for a class of flows [5], a KAM-type theorem exists [6], and with it the associated barriers to global transport [4]. On the other hand, the case of two actions, which is also generic for a class of flows, displays in Liouvillian maps a new phenomenon of resonance-induced diffusion leading to global transport throughout phase space [4]. However, despite this theoretical progress, until now no realistic flow — in the sense of being at least an approximate solution of the Navier–Stokes equations for a realizable experiment — has been observed to show the above properties. The purpose of this Letter is to fill the gap for the two-action case by introducing the first engineerable flow demonstrating the presence of resonance-induced diffusion in real fluid flows.

We begin by briefly illustrating the phenomenon for a map, introducing at the same time the adiabatic invariants and resonant surfaces we encounter in two-action flows. The map

$$I_1' = I_1 + \varepsilon F_1(I_2, \theta), \qquad I_2' = I_2 + \varepsilon F_2(I_1', \theta),$$
$$\theta' = \theta + \omega(I_1', I_2'), \qquad (1)$$

represents a small perturbation of an integrable Liouvillian map having two action and one angle variables [7]: for vanishing $\varepsilon$, the two action variables $I_1$ and $I_2$ remain invariant while the angle $\theta$ rotates with a constant angular frequency that depends only on the actions. When $\varepsilon$ is nonzero but small the action variables drift slowly compared to the angle. This separation of scales allows us to average over $\theta$ leading to an adiabatic description of the motion. Before the actions change appreciably, the angle is able to traverse a large sample of its domain. Hence the evolution of the actions is sensitive only to the average value of the angle: the action equations can be averaged over the angle, and so become decoupled from the angle equation to yield an area-preserving map

$$I_1' = I_1 + \varepsilon \bar{F}_1(I_2), \qquad I_2' = I_2 + \varepsilon \bar{F}_2(I_1'), \qquad (2)$$

where the bar represents the $\theta$ average. Being a small perturbation of the two-dimensional identity, the map of the action plane can be well approximated by a two-dimensional autonomous Hamiltonian flow

$$\dot{I}_1 = \bar{F}_1(I_2) = \frac{\partial H}{\partial I_2}, \qquad \dot{I}_2 = \bar{F}_2(I_1) = -\frac{\partial H}{\partial I_1}. \qquad (3)$$

One is then led to the conclusion that the dynamics of the map in Eq. (1) occurs on invariant surfaces that are the product of the almost uniform angular motion of $\theta$, and the level curves of the Hamiltonian

$$H(I_1, I_2) = \int_0^{I_2} \bar{F}_1(I) dI - \int_0^{I_1} \bar{F}_2(I) dI = \beta. \qquad (4)$$

However, this analysis breaks down when the angular variable does not sample its domain of variation uniformly. This failure is bound to occur when the rotation is resonant; when $n\omega(I_1, I_2) = 2\pi k$, where $k$ and $n$ are integers. The family of curves in the action plane defined by Eq. (4) is generically transversal to the curves on which these resonances occur. Although the resonances are dense in the action plane, they have a hierarchical structure in which the relative strength of different orders $k/n$ of resonances is governed by the Fourier expansion of Eq. (1), such that in general the lowest-order resonances have the smallest denominators $n$. Hence one might expect the adiabatic approximation to provide a good descrip-tion of the action motion, except at the intersections with the lowest-order resonances. This is depicted in Fig. 1(a) for a particular two-action Liouvillian map. Close to resonances the trajectory oscillates wildly between different invariant curves $H = \beta$. In Fig. 1(b) we have plotted the time evolution of the adiabatic invariant $H$ for this map to show that it remains nearly constant almost all the time, but jumps chaotically from one invariant curve to another at each intersection.

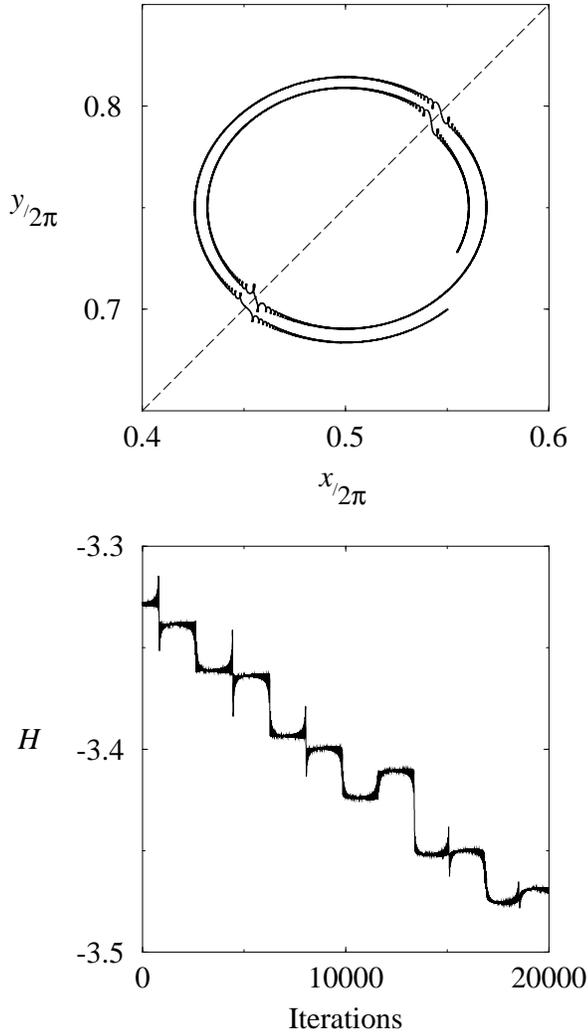

FIG. 1. (a) Projection of a trajectory in a two-action Liouvillian map onto the plane $z = 0$. The map is $x' = x + 0.001(\sin z + 2\cos y)$, $y' = y + 0.001(1.5\sin x' + 2.5\cos z)$, $z' = z + 4(\cos y' + \sin x')$. The diagonal dashed line shows the location of the lowest-order 0/1 resonance. (b) The time evolution of the corresponding adiabatic invariant $H = 2\sin y + 1.5\cos x$.

The study of two-action properties in a real fluid flow is made difficult by the fact that very few three-dimensional time-dependent fluid flows are analytically tractable. In the following we introduce a fluid flow that can be solved analytically as a perturbation expansion of the Navier–Stokes equations and that exhibits the phenomenon just illustrated.

The flow we have chosen to consider is incompressible flow at low Reynolds numbers between two concentric spheres which rotate with different angular velocities about a common axis that switches in turn between two different directions separated by an angle $\alpha$. In the absence of time dependence introduced by the axis switching, the flow consists of a primary spherical Couette flow about the rotation axis, superposed with an orthogonal secondary flow in the meridian plane $(r, \theta)$ [8]. Owing to axial symmetry, none of the flow components depend on the azimuthal coordinate $\phi$. The secondary flow, which is made up of one or two Taylor vortices in each hemisphere, depending on the flow parameters [9], is then two dimensional and described by a stream function $\psi$ that can be computed perturbatively in the Reynolds number $Re$. The velocity field up to first order in the Reynolds number reads

$$v_r = \frac{1}{r^2 \sin\theta} \frac{\partial \psi}{\partial \theta} = \dot{r}, \qquad v_\theta = -\frac{1}{r \sin\theta} \frac{\partial \psi}{\partial r} = r\dot{\theta},$$
$$v_\phi = \left(a_1 r + \frac{a_2}{r^2}\right) \sin\theta = r \sin\theta \, \dot{\phi}, \qquad (5)$$

where $\psi$ is given by

$$\psi = Re\Big(\frac{A_1}{r^2} + A_2 + A_3 r^3 + A_4 r^5 + \frac{a_2}{4}\left(\frac{a_2}{r} - a_1 r^2\right)\Big) \sin^2\theta \cos\theta, \qquad (6)$$

and $a_1$, $a_2$, $A_1$, $A_2$, $A_3$, $A_4$, are constants dependent on the ratios of the radii and the angular velocities of the spheres [9]. Notice that for $Re \ll 1$ the secondary flow $(v_r, v_\theta)$ is very much slower than the primary flow $v_\phi$.

The perturbation of this flow by periodic axis switching introduces time dependence at the same time as true three dimensionality by coupling together the velocity components of Eq. (5). Since $Re \ll 1$, we can assume that fluid inertia is not important, such that trajectories of passive scalars will be a piecewise juxtaposition of the steady flow about each axis. Notice that the $r$ and $\theta$ coordinates are always coupled by the secondary flow except when the Reynolds number is precisely zero. Furthermore, adding a second rotation axis leaves the equation describing the evolution of $r$ unchanged, but introduces a coupling between $\theta$ and $\phi$ which is of $O(\alpha)$ for small axis separation $\alpha$. All this implies that, although we do not have direct access to the three-dimensional integrable case, we can however intuit that it should exist close by the case we are considering for sufficiently small perturbations. We can construct a reference frame $(r, \tilde\theta, \tilde\phi)$ tilted midway between the two axes [9], with the rationale that each semiperiod has its own adiabatic invariant, Eq. (6), tilted with respect to the other, and we expect that the composed trajectory will on average lie on the adiabatic invariant midway between the two. We can show that this will be the case in the limit when the axis separation and the Reynolds number are sufficiently small. The change in $r$ and $\tilde\theta$ over a period is bounded by an quantity of $O(\alpha)$, while $\tilde\phi$ changes by an arbitrary amount over the same period. Thus for small axis separation $\alpha$ and low Reynolds numbers, $r$ and $\tilde\theta$ are action variables, and $\tilde\phi$ is an angle. Having this set of approximate action–angle variables, we are able to calculate the invariant surfaces and the resonant surfaces associated with a nonintegrable perturbation of a two-action flow.

adiabatic invariant and the resonances for a particular value of the parameters.

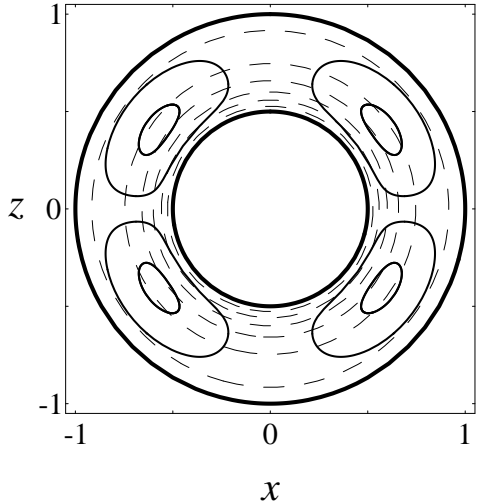

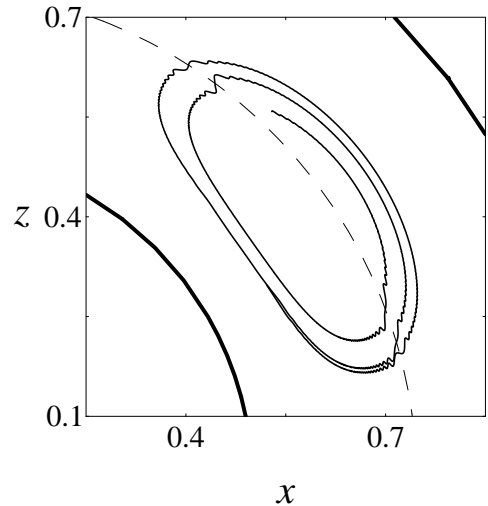

FIG. 2. Level curves of the adiabatic invariant surfaces (thin lines) are shown inside the two spheres (thick lines) together with the lowest-order resonant surfaces intersecting them (thin dashed lines) on the plane $y = r \sin\theta \sin\phi = 0$ for a particular set of parameter values for small axis separation. The resonances shown are, from the inner sphere to the outer, $k/n = -2/1, -3/2, -1/1, -1/2, 0/1$, and $1/2$.

Although we are not able to obtain an exact expression for the adiabatic invariant of the stroboscopic map for all $\alpha$, we can obtain perturbatively in $\alpha$ an expression

$$H = \frac{Re}{2}\left(\frac{A_{11}}{r^2} + A_{21} + A_{31}r^3 + A_{41}r^5 \right.$$
$$+ \frac{a_{21}}{4}\left(\frac{a_{21}}{r} - a_{11}r^2\right) + \frac{A_{12}}{r^2} + A_{22} + A_{32}r^3 + A_{42}r^5$$
$$\left. + \frac{a_{22}}{4}\left(\frac{a_{22}}{r} - a_{12}r^2\right)\right) \sin^2\tilde{\theta}\cos\tilde{\theta} \qquad (7)$$

valid for small axis separations, where $a_{ij}$, and $A_{ij}$ represent the constants $a_i$ and $A_i$ evaluated for the two semiperiods $j = 1, 2$. We can similarly obtain the resonant surfaces: in the tilted coordinate system $(r, \tilde{\theta}, \tilde{\phi})$, $\tilde{\phi}$ correspondingly strobed is the fast angular variable. The resonant rotation of $\tilde{\phi}$, as in Liouvillian maps, occurs when $\Delta\tilde{\phi} = 2\pi k/n$, where $k$ and $n$ are integers. For small axis separations, this is true when

$$\Delta\tilde{\phi} = \frac{2\pi k}{n} = \frac{1}{2}\left(a_{11} + \frac{a_{21}}{r^3} + a_{12} + \frac{a_{22}}{r^3}\right) T, \qquad (8)$$

where $T$ is the axis-switching period. Notice that these resonant surfaces are independent of $\theta$, i.e., are spherical shells of varying radii $r_{k/n}$, up to first order in the axis separation. In Fig. 2 we show the locations in the meridian plane of both the

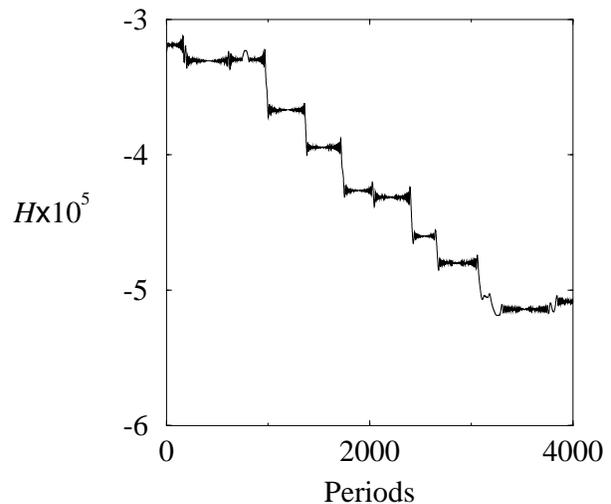

FIG. 3. (a) The projection of a strobed trajectory onto the plane $y = r \sin\theta \sin\phi = 0$. The parameter values were chosen to be the same as in Fig. 2, such that the lowest-order $0/1$ resonance (dashed) is in the middle of the region between the spheres, while leaving other primary resonances distant. The axis separation angle is $\alpha = 0.1°$ and the Reynolds number $Re = 0.1$. (b) Time evolution of the adiabatic invariant $H$ for this case.

Numerically computed strobed trajectories show that the theoretical picture we painted above is accurate and that the phenomenon of resonance-induced diffusion, previously reported only for Liouvillian maps, is present here with striking resemblance in a real fluid flow. To illustrate this, we choose the flow parameters such that the relative geometry of the adiabatic invariants and the resonances matches that of Fig. 1(a) above: the lowest-order resonance in Fig. 2 passes near the middle of the family of adiabatic invariants. In Fig. 3(a) we show a projection on the meridian plane of the stroboscopically-sampled trajectory of the flow at these parameter values. The similarity between Fig. 3(a) and Fig. 1(a) is immediately obvious. Furthermore, in Fig. 3(b), we show the strobed time evolution of the invariant given by Eq. (7). Again, its likeness to Fig. 1(b) is undeniable, which highlights both the presence of resonance-induced diffusion here, and the accuracy of our assumptions about the adiabatic invariant.

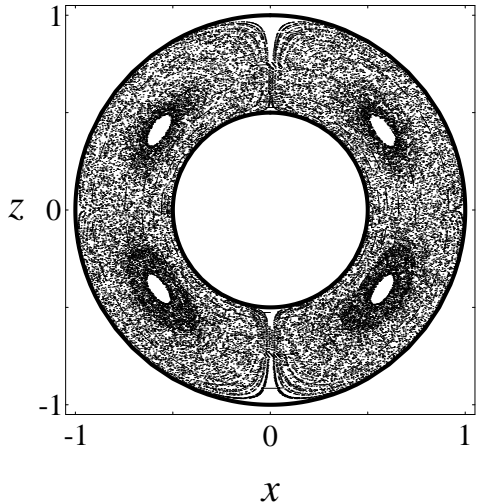

FIG. 4. A slice of a strobed trajectory between $-0.01 < y < 0.01$ for the same parameter values as Fig. 3. Locations of low-order resonances are shown as in Fig. 2. Forty thousand points (corresponding to many times this number of periods) are plotted here, all from the same initial condition.

Finally, as the slice of a strobed trajectory in Fig. 4 attests, the mechanism of resonance-induced diffusion is responsible for the existence of asymptotically-globally-space-filling trajectories. This slice illustrates how effective resonance-induced diffusion is at mixing; the perturbation, which is of $O(\alpha)$, is very small here, yet the only regions where the trajectory has not yet ventured are close to the poles, and the middle of the vortices, which lie inside the $0/1$ resonance shell. In time, the trajectory will diffuse into these regions too, through the action of the higher-order resonances that are dense throughout the space. Increasing the axis separation $\alpha$ increases the amount of time the trajectory is captured into resonance, and with it the size of the jumps and the diffusion rate. Other quantities relevant to mixing — the stretching and folding measured by the Lyapunov exponents of the flow — also increase as powers of $\alpha$. Experimental control of the mixing rate can then be achieved through adjustment of the strength and density of the resonances with the axis separation $\alpha$, the angular velocity ratio of the spheres, and the period of the motion [9]. Further quantitative studies of mixing efficiency, as well as the effects of the interplay of space-filling trajectories and molecular diffusion on mixing and transport, are now in progress.

In summary, we have shown that the properties of nearly-integrable two-action Liouvillian maps are highly relevant to the transport features of a large class of real fluid flows. We expect to see in the stroboscopic maps of such flows the resonance-induced diffusion characteristic of two-action maps, consisting of motion on invariant surfaces interspersed with periods of motion on resonant surfaces. We have also shown that our action–angle classification of Liouvillian maps has good predictive capability. This finding opens up an avenue for the experimental verification of the existence of space-filling trajectories in this and other similar flow geometries in the presence of small periodic modulations. There are many technological implications of this novel diffusion phenomenon, and we aim to have excited the reader about the new possibilities for enhanced mixing that may arise.

MF acknowledges the support of the Israel Science Foundation administered by the Israel Academy of Sciences and Humanities, and OP and JHEC that of the Spanish Dirección General de Investigación Científica y Técnica, contract numbers PB94-1167 and PB94-1172 and European Union Human Capital and Mobility contract number ERBCHBICT920200.

[*] Email julyan@hp1.uib.es, Web http://formentor.uib.es/~julyan.
[†] Email mario@bgumail.bgu.ac.il.
[‡] Email piro@hp1.uib.es, Web http://formentor.uib.es/~piro.
[1] Note that, following nonlinear dynamics terminology, we talk here of *diffusion*, whereas in fluid dynamics one would rather talk of *dispersion*.
[2] See, for example, J. M. Ottino, *The Kinematics of Mixing: Stretching, Chaos, and Transport* (Cambridge University Press, Cambridge, 1989); Annu. Rev. Fluid Mech. **22** 207 (1990); *Chaotic Advection, Tracer Dynamics and Turbulent Dispersion*, edited by A. Babiano, A. Provenzale, and A. Vulpiani (North-Holland, Amsterdam, 1994).
[3] See, for example, T. Dombre, U. Frisch, J. M. Greene, M. Hénon, A. Mehr, and A. M. Soward, J. Fluid Mech. **167**, 353 (1986); D. V. Khakhar, H. Rising, and J. M. Ottino, J. Fluid Mech. **172**, 419 (1986); S. W. Jones, O. M. Thomas, and H. Aref, J. Fluid Mech. **209**, 335 (1989); H. A. Kusch and J. M. Ottino, J. Fluid Mech. **236**, 319 (1992); R. S. MacKay, J. Nonlinear Sci. **4**, 329 (1994). Y. Le Guer, C. Castelain, and H. Peerhossaini, J. Fluid Mech. (to be published);
[4] O. Piro and M. Feingold, Phys. Rev. Lett. **61**, 1799 (1988); M. Feingold, L. P. Kadanoff, and O. Piro, J. Stat. Phys. **50**, 529 (1988); J. H. E. Cartwright, M. Feingold, and O. Piro, Physica D **76**, 22 (1994).
[5] I. Mezić and S. Wiggins, J. Nonlinear Sci. **4**, 157 (1994).
[6] C.-Q. Cheng and Y.-S. Sun, Celest. Mech. **47**, 275 (1990).
[7] While this is not the most general form for such a map, it represents a subclass that is explicitly iterable.
[8] B. R. Munson and D. D. Joseph, J. Fluid Mech. **49**, 289 (1971).
[9] J. H. E. Cartwright, M. Feingold, and O. Piro, J. Fluid Mech. (to be published).